\documentclass{elsart}[12pt] 
\usepackage{graphicx}

\usepackage{amssymb}

\begin{document}

\begin{frontmatter}


\title{Electronic, magnetic, and vibrational properties of the molecular
magnet Mn$_4$ monomer and dimer}


\author[1,2]{Kyungwha Park}
\author[1]{Mark R. Pederson\corauthref{cor1}}
\corauth[cor1]{Corresponding Author: Tel.+1 202 767 3413; Fax. +1 202 404 7546;
E-mail:pederson@dave.nrl.navy.mil}
\author[1]{Noam Bernstein}
\address[1]{Center for Computational Materials Science, Code 6390,
Naval Research Laboratory, Washington DC 20375}
\address[2]{Department of Electrical Engineering and Materials
Science Research Center, Howard University, Washington DC 20059}

\begin{abstract}
A new type of the single-molecule magnet
[Mn$_4$O$_3$Cl$_4$(O$_2$CEt)$_3$(py)$_3$]
forms dimers. Recent magnetic hysteresis measurements on this single-molecular
magnet revealed interesting phenomena: an absence of quantum tunneling 
at zero magnetic field and tunneling before
magnetic field reversal. This is attributed to a significant antiferromagnetic
exchange interaction between different monomers. To investigate this system,
we calculate the electronic structure, magnetic properties, intramolecular and
intermolecular exchange interactions using
density-functional theory within the generalized-gradient approximation.
Our calculations agree with experiment. We also calculate vibrational
infrared absorption and Raman scattering intensities for the monomer
which can be tested experimentally.
\end{abstract}

\end{frontmatter}


\section{Introduction}

Single-molecule magnets (SMMs) are three-dimensional arrays of
identical nanoscale molecules, each of which consists of several
transition metal ions surrounded by organic ligands. These SMMs
are a promising building block for magnetic storage devices and for
investigating macroscopic quantum tunneling\cite{CHUD98} and
quantum decoherence.\cite{STAM03} A prototype of these SMMs is
[Mn$_{12}$O$_{12}$ \linebreak
(CH$_3$COO)$_{16}$(H$_2$O)$_4$]$\cdot$2(CH$_3$COOH)$\cdot$4(H$_2$O)
(hereafter Mn$_{12}$)\cite{LIS80} which has an effective ground-state
spin of $S=10$. Magnetic hysteresis measurements on the SMM
Mn$_{12}$ showed quantum tunneling between spin-up states and
spin-down states. \cite{SESS93,FRIE96,BARR97,PERE98} This has
been successfully explained by magnetic anisotropy parameters
only. However, recent measurements on a new type of SMM,
[Mn$_4$O$_3$Cl$_4$(O$_2$CEt)$_3$(py)$_3$] (hereafter
Mn$_4$)\cite{HEND92,EDWA03} where Et=CH$_2$CH$_3$ and
py=NC$_5$H$_5$=pyridine, showed qualitatively different tunneling
behavior\cite{WERN02-NAT}: quantum tunneling occurred before the
magnetic field was reversed and no quantum tunneling was observed
at zero field, which is in contrast to other SMMs like Mn$_{12}$.
It was speculated that this behavior was caused by substantial
antiferromagnetic exchange interactions between different
monomers within a dimer.\cite{WERN02-NAT} Suppose that magnetic
moments of all molecules of the SMM Mn$_4$ are aligned due to
strong external magnetic field. If the direction of the external
magnetic field is reversed adiabatically through zero field, then
some molecules may change their magnetic moments before the
external field reversal. This is because the antiferromagnetic
exchange interactions facilitate monomeric magnetic moment flips
prior to the external field reversal. This interaction
also prevents two monomers from simultaneously flipping their
magnetic moments at zero field. In contrast, it is widely accepted 
that for Mn$_{12}$ there are no appreciable exchange interactions between
different monomers. In this sense, further investigation of the SMM
Mn$_4$ is worthwhile. Additionally, the SMM Mn$_4$ consists of
the same kinds of Mn spins as those for Mn$_{12}$ but it is much
simpler than Mn$_{12}$ since the number of Mn spins is smaller.
Therefore the SMM Mn$_4$ can be used to obtain some insights into
issues remaining in the SMM Mn$_{12}$ such as clarifying the origins
of symmetry-breaking terms in the effective single-spin
Hamiltonian\cite{CHUD01,CORN02} and possible spin-orbit-vibron
interactions.\cite{PEDE02}

\begin{figure}
\begin{center}
\raisebox{.25\textwidth}{
\includegraphics[angle=0,width=0.2\textwidth]{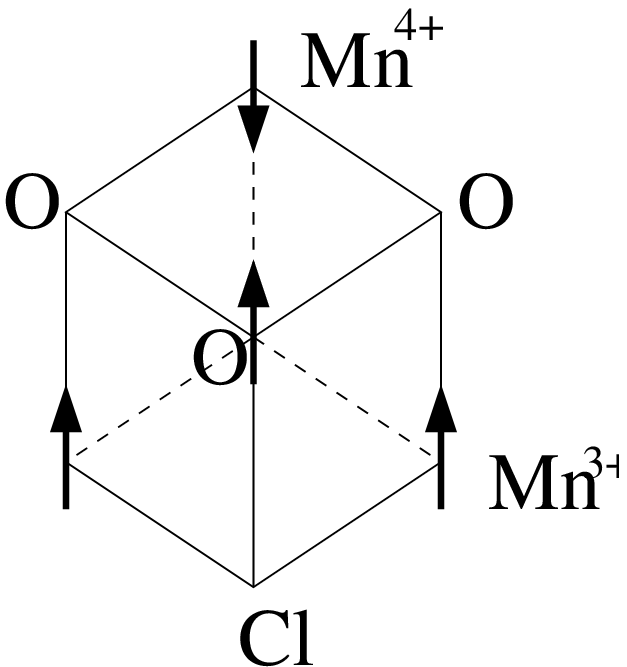}}
\vspace{.2in}
\includegraphics[angle=0,width=0.35\textwidth]{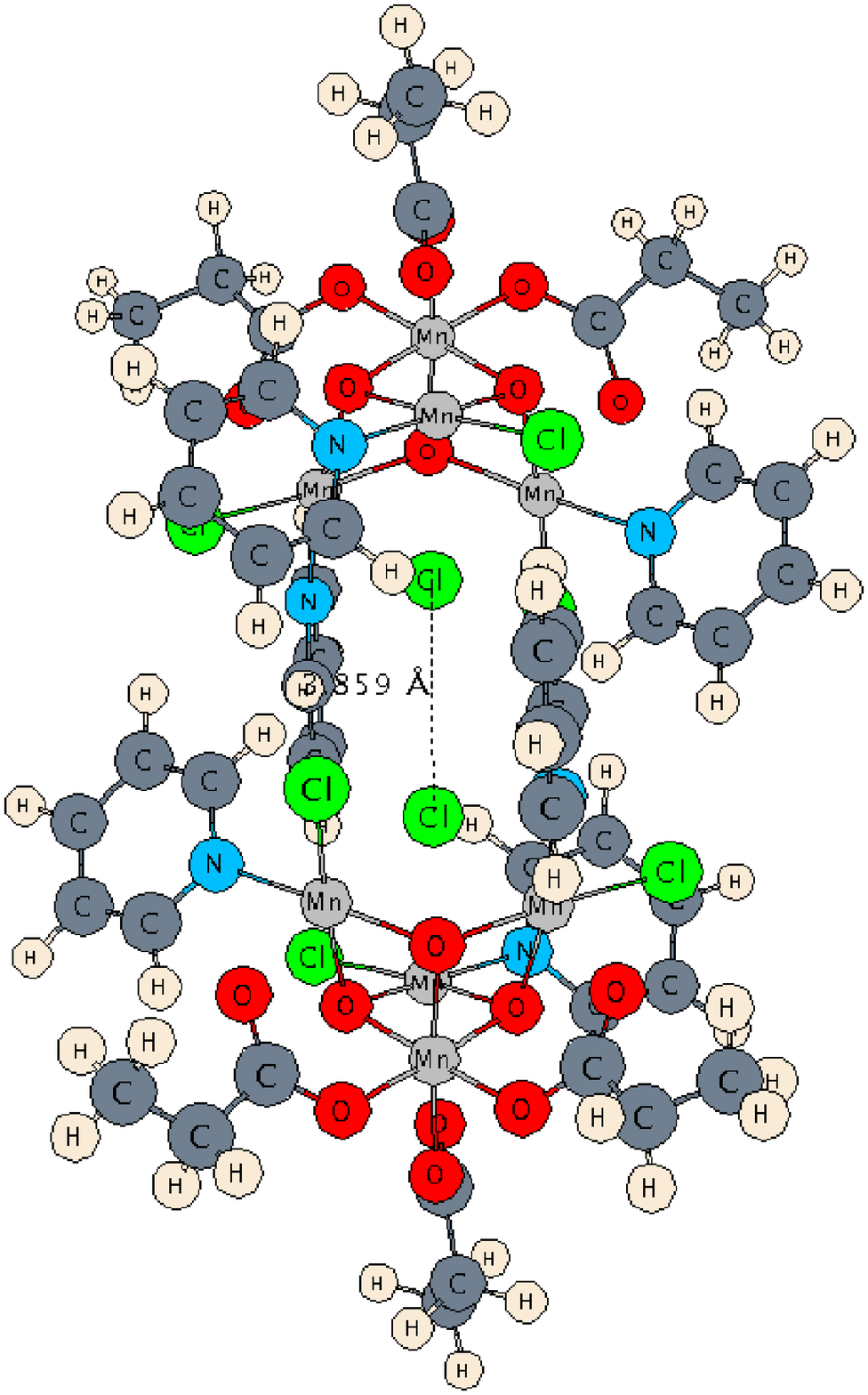}
\caption[]{(Left) Cubane-like magnetic core of the Mn$_4$ monomer.
The four arrows represent Mn spins and O and Cl atoms are labeled.
The magnetic core consists of three ferromagnetically coupled
Mn$^{3+}$ spins ($S$$=$2) coupled antiferromagnetically to
one Mn$^{4+}$ spin ($S$$=$3/2) ion leading to a total spin of $S=9/2$.
(Right) Mn$_4$ dimer geometry. The dimer is formed by inversion
of the threefold symmetric monomer. The distance between
the two central Cl atoms marked as the dotted line was measured
to be 3.86~\AA.~The threefold axis is along this Cl-Cl bond. }
\label{fig:Mn4geo}
\end{center}
\end{figure}

A dimeric form of the SMM Mn$_4$ is obtained by inversion symmetry
of the threefold symmetric monomer shown in Fig.~\ref{fig:Mn4geo}.
Since dimers are well separated from each other, interactions
between different dimers are negligible so that they are not
considered in our calculations. The magnetic core of the Mn$_4$
monomer comprises three ferromagnetically coupled Mn$^{3+}$
($S=2$) ions coupled antiferromagnetically to the remaining
Mn$^{4+}$ ($S=3/2$) ion leading to a total ground-state spin of
$S=2\times 3 - 3/2 \times 1=9/2$. The core has a similar cubane
structure to the inner core of the SMM Mn$_{12}$, although in Mn$_{12}$
there are four Mn$^{4+}$ ions. In this work, we investigate the
electronic structure, magnetic properties, and vibrational modes
of the SMM Mn$_4$ using density-functional theory
(DFT).\cite{KOHN65} In Sec.~2.1, we present calculations of the
Mn$_4$ monomer such as optimized geometries, intramolecular
exchange interactions between Mn spins, and magnetic anisotropy
barriers for the ground-state and excited-state manifolds. In
Sec.~2.2, we discuss binding energy and exchange interactions
between different monomers within a dimer. In Sec.~2.3, we
present calculated infrared and Raman intensities for vibrational
normal modes. In Sec.~3, we present our conclusion.

\section{DFT calculations and Discussion}

We have used all-electron Gaussian-orbital-based Naval Research
Laboratory Molecular Orbital Library (NRLMOL)
\cite{PEDE90,JACK90,PEDE91,BRIL98,QUON93,PORE97,PORE96} within
the Perdew-Burke-Ernzerhof (PBE) generalized gradient
approximation (GGA)\cite{PERD96} along with full basis sets for
all atoms and fine mesh.\cite{PORE99} In our analysis, we have
considered monomers and dimers that are terminated by both H and
by the CH$_2$CH$_3$ radicals found in the experiments. We have
constructed a dimer comprising two conformers of the monomeric
units. Our study showed that both conformers are stable. Our
vibrational analysis on a monomer terminated by H confirms that
there are no unstable modes. The two conformers have slightly
different arrangements of the pyridine ligands. The first
conformer was found by the density-functional-based optimization
of the hydrogenated geometry. It is called the computationally
determined conformer (CDC). The second conformer was obtained by
the experimental x-ray data with hydrogen positions corrected.
This is called the experimentally determined conformer (EDC).

\subsection{Monomeric electronic and magnetic properties}

Details of finding the CDC monomer were discussed
elsewhere\cite{PARK03} so here they are briefly summarized. The
Mn$_4$ monomer has threefold symmetry about the direction of the
central bridging Cl-Cl bond in the dimer geometry shown in
Fig.~\ref{fig:Mn4geo} (or the direction of connecting
Mn$^{4+}$ and Cl in the cubane). To find an initial geometry,
we optimize a pyridine ring and a cubane separately. Then the
initial geometry for the monomer is relaxed using NRLMOL with the
Cl atom fixed to reproduce the experimental Cl-Cl distance upon
dimerization. Relaxation continues until maximum forces between
atoms are no larger than $\sim 0.001$~hartree/bohr. A geometry
for the dimer can be found by inversion symmetry of the optimized
monomer geometry. Charges and magnetic moments for the Mn atoms
from the CDC monomer agree well with those from the EDC monomer.
The total magnetic moment is 9$\mu_B$, which agrees with experiment.
The energy gap between the minority lowest unoccupied molecular
orbital (LUMO) and the majority highest occupied molecular orbital
(HOMO) is much larger than the thermal energy shown in
Table~\ref{table:1}, so the total magnetic
moment of the ground state is stable. Calculated electronic
density of states (not shown) confirms that the three Mn$^{3+}$
spins are antiferromagnetically coupled to a Mn$^{4+}$ spin.

\begin{figure}
\begin{center}
\includegraphics[angle=0,width=0.3\textwidth]{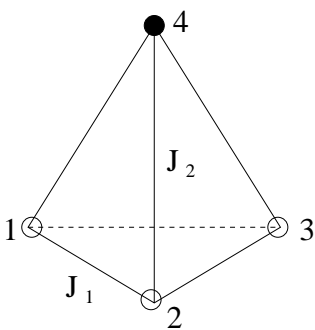}
\caption{Schematic diagram of intramolecular exchange interactions within a monomer.
Three Mn$^{3+}$ ($S=2$) spins are denoted as empty circles and one
Mn$^{4+}$ ($S=3/2$) spin as a filled circle at the four vertices of the tetrahedra.
Each spin is numerically labeled at each vertex. $J_1$ is the exchange interaction between
Mn$^{3+}$ spins and $J_2$ is between Mn$^{3+}$ and Mn$^{4+}$ spins.}
\label{fig:spin_conf}
\end{center}
\end{figure}

Since both the CDC and EDC monomers are stable, we choose one of them,
for example, the optimized CDC monomer to examine intramolecular
exchange interactions. Considering the symmetry of the monomer
and the fact that there are two kinds of Mn spins, we recognize
that there are two types of exchange interactions between Mn spins
within a monomer as shown in Fig.~\ref{fig:spin_conf}. To
calculate the exchange coupling constants, we assume that
magnetic moments of all Mn spins are aligned along a particular
direction and of Ising type. So a Mn$^{3+}$ spin can have either
$M_z=+2$ or $M_z=-2$ and a Mn$^{4+}$ spin can have $M_z=+3/2$ or
$M_z=-3/2$. Thus we can construct three different spin
configurations other than the ground-state shown in
Table~\ref{table:2}. For example, $M_s=15/2$ ($M_s$ is an
eigenvalue of the $z$ component of the total spin operator $S$)
is achieved by flipping a Mn$^{4+}$ spin from the ground-state
$S=9/2$: $M_s=2+2+2+3/2=15/2$. Notice that all of the examined
spin configurations are not eigenstates of $S^2$. We set up
geometries of the three spin configurations using the
optimized ground-state geometry with corresponding Ising-type spin
arrangements and proper magnetic moments. Then we minimize
self-consistently the energies of the spin configurations.
Comparing the energies of the three configurations (refer to
Table~\ref{table:2}) with the energy gap between the minority
LUMO and the majority HOMO, we find that the examined spin
configurations have an order of magnitude smaller energy than
other kinds of spin excitations such as moving one majority spin
to the unoccupied minority orbital. So we call those spin
configurations low-energy spin excitations. Since there are three
unknowns (the background energy, $E_0$, the two types of exchange
constants, $J_1$ and $J_2$) and four equations to solve, we can
calculate the exchange constants by the least-square-fit (LSF)
method.  Our calculated values of $J_1$ and $J_2$ are presented
in Table~\ref{table:3}.

\begin{figure}
\begin{center}
\includegraphics[angle=0,width=0.4\textwidth]{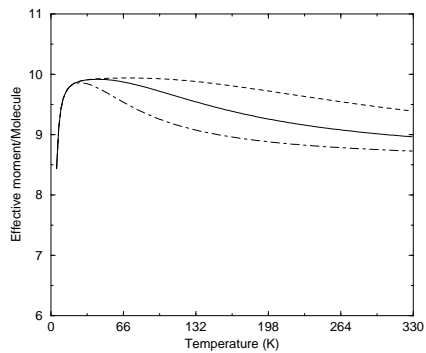}
\caption{Calculated effective moment $\mu_{\mathrm{eff}}$ per monomer.
The dashed, solid, and dot-dashed curves are obtained using the calculated
values of $J_1$ and $J_2$, a half of the calculated values, and
a quarter of the calculated values respectively. The solid line is close to
the experimental data.\cite{HEND92}}
\label{fig:mag_suscep}
\end{center}
\end{figure}

Differences between the calculated energies and
the LSF-determined values for the ground state and
low-energy spin excitations are quite small as shown in
Table~\ref{table:2}. The exchange constant between
Mn$^{3+}$ spins, $J_1$, is confirmed to
be ferromagnetic, and the exchange constant between
Mn$^{3+}$ spins and a Mn$^{4+}$ spin, $J_2$, is antiferromagnetic
but larger than $J_1$. From exact diagonalization of
Heisenberg exchange Hamiltonian, we confirm that the ground-state
manifold has $S=9/2$, that the first excited-state manifolds
are doubly degenerate with $S=7/2$, and that the second excited-state
manifold has $S=11/2$. The energy gaps between the
ground-state manifold and the excited-state manifolds
are given in Table~\ref{table:3}. The large energy gap between the
ground-state and the first-excited manifold also supports that the
ground-state $S=9/2$ manifold is stable. To check how good the
calculated values of $J_1$ and $J_2$ are compared to experiment,
we calculate the effective moment per molecule, $\mu_{\mathrm{eff}}$,
at magnetic field of 1~T as a function of temperature $T$ as
follows:
\begin{eqnarray}
\mu_{\mathrm{eff}}=\frac{\sqrt{3 \chi k_B T}}{\mu_B}
\end{eqnarray}
where $\chi$ is susceptibility, $k_B$ is the Boltzmann constant,
and $\mu_B$ is the Bohr magneton. If we use the calculated values
of $J_1$ and $J_2$ given in DFT(1) of Table~\ref{table:3}, then
we obtain overestimated effective moments (dashed curve in
Fig.~\ref{fig:mag_suscep}) at high temperatures compared to the
experimental data. If we use half-reduced values of $J_1$ and
$J_2$ from the DFT-calculated ones, then the calculated effective
moment (solid curve in Fig.~\ref{fig:mag_suscep}) agrees well
with experiment. Notice that the reduced values of $J_1$ and $J_2$
are quite close to the experimentally extracted values in
Table~\ref{table:3}. This supports the argument that DFT
calculations often overestimate the exchange interactions between
atoms because of imperfect treatment of self-interaction of the
Coulomb potential.

We calculate the magnetic anisotropy barrier (MAE) in zero
magnetic field for both the CDC and EDC monomers for the ground
state $S=9/2$, assuming that the spin-orbit coupling mainly
contributes to the barrier. Following the procedure explained in
Ref.~\cite{PEDE99}, we find that the Mn$_4$ monomer has the
easy axis along the threefold axis, in agreement with
experiment. The calculated anisotropy barrier is about 11.3~K for
the CDC monomer, 11.6~K for the EDC monomer, and 10.9~K for the
hydrogenated EDC monomer. All of these values are close to the
experimentally measured value of 14.4~K. To investigate the
contribution of each Mn spin to the barrier, we project the total
barrier onto each Mn spin and calculate its projected anisotropy.
The three Mn$^{3+}$ spins have the easy axes along the $x$, $y$,
and $z$ axis respectively. The Mn$^{4+}$ spin has the easy axis
along the $\langle 111 \rangle$ direction. Most of the anisotropy originates
from the Mn$^{3+}$ spins due to Jahn-Teller distortion,
while the contribution of the Mn$^{4+}$ spin is insignificant.
The same tendency occurs for the SMM Mn$_{12}$. We also calculate
the magnetic anisotropy barriers for the low-energy spin
excitations shown in Table~\ref{table:2}. The barriers are
essentially the same as that for the ground state $S=9/2$.

\subsection{Dimeric properties}

For a dimeric form, the distance between the central bridging
Cl-Cl bond is kept as the experimental value of 3.86~\AA~unless
specified. We calculate the binding energy of the dimer by
subtracting the dimer energy from twice the monomer energy.
We find that the dimer is stable for both the CDC and EDC.
For the CDC (hydrogenated EDC) dimer, the binding energy is about
0.16~eV (0.45~eV).\cite{PARK03} The magnitude of the binding
energy suggests attractive electrostatic interactions between
different monomers. We find that the electric dipole moment for
each CDC (hydrogenated EDC) monomer is as large as 2.28 (1.91) 
in atomic units and that it points towards the Cl atom in the 
cubane along the threefold axis.
Since the dimer has inversion symmetry, the electric dipole moment
for the dimer vanishes.
The discrepancy between the binding energy for the CDC and
that for the EDC arises from our substitution of CH$_2$CH$_3$
for H and slightly different geometries of pyridine rings for both
conformers.

To calculate the exchange coupling constant $J_{\mathrm{inter}}$
between different monomers within a dimer, we make assumptions
that a monomer is an ideal $S=9/2$ object and that its effective
spin is aligned along the easy axis and of Ising type
(either $M_z=+9/2$ or $M_z=-9/2$).
Then we calculate self-consistently the energies of ferromagnetic
(parallel monomeric spins) and antiferromagnetic (antiparallel
monomeric spins) configurations of the dimer, and take the
difference $\delta$ between the two energies. We find that the
antiferromagnetic configuration is favored, and that from
$\delta=2J_{\mathrm{inter}}(9/2)^2$ the exchange constant
$J_{\mathrm{inter}}$ is about 0.24~K for the CDC and 0.27~K
for the EDC.\cite{PARK03} This can be compared to the experimentally
measured value of 0.1~K. From the analysis of the exchange
constants within a monomer, this overestimated value of
$J_{\mathrm{inter}}$ is not so surprising.
We also find that $J_{\mathrm{inter}}$ is quite sensitive to
the central bridging Cl-Cl bond length.
Our calculations show that
$J_{\mathrm{inter}}$ increases exponentially with
decreasing the Cl-Cl distance. More detailed analysis was given in
Ref.\cite{PARK03}. Since the experimental results cannot
distinguish exchange interaction from magnetic dipole-dipole
interaction, we estimate the contribution of the magnetic
dipole-dipole interaction to the total interactions between
different monomers in a dimeric form. Since the dipole moment
is along the easy axis and the two monomers are located along
the easy axis, the ferromagnetic configuration is favored.
But the difference in the dipole-dipole interaction between the
ferromagnetic and antiferromagnetic configuration is about 0.28~K,
which is negligible compared to the exchange energy difference of
10~K. This was calculated considering two large spins of $S=9/2$.
Including the internal structure of a monomer does not
substantially change the energy difference in the dipole-dipole interaction.

\subsection{Vibrational analysis}

\begin{figure}
\begin{center}
\includegraphics[angle=0,width=0.6\textwidth]{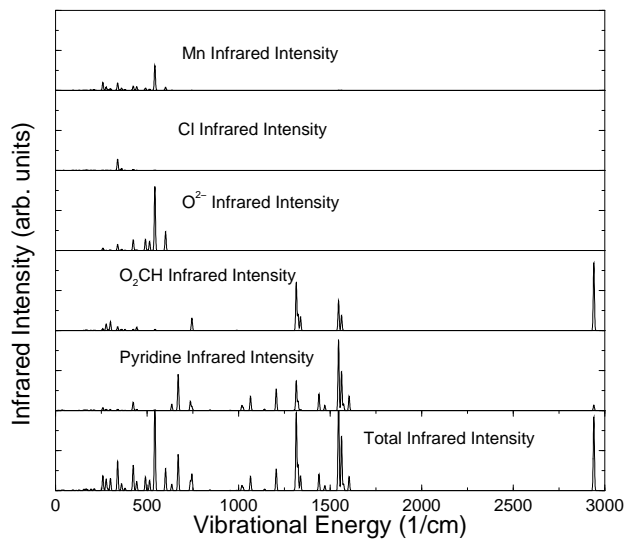}
\caption{Calculated total and projected infrared (IR) vibrational spectra
for the experimentally determined conformer (EDC) with CH$_2$CH$_3$ replaced by H.
From the top panel, shown are the projected IR active density of
states onto Mn, Cl, O$^{2-}$, O$_2$CH, and pyridine ligands, and
the total IR density of states. All of the projected IR density of states
have the same scale as the total IR density of states.}
\label{fig:IR}
\end{center}
\end{figure}

\begin{figure}
\begin{center}
\includegraphics[angle=0,width=0.6\textwidth]{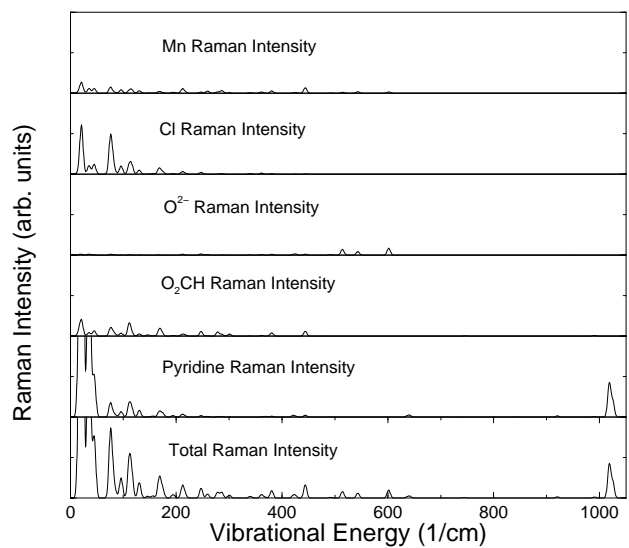}
\caption{Calculated total and projected Raman vibrational spectra
for the same monomer used for the IR spectra. All of the
projected Raman intensities have the same scale as the total
Raman intensity.}
\label{fig:Raman}
\end{center}
\end{figure}

We calculate vibrational normal modes for the EDC monomer with
CH$_2$CH$_3$ substituted by H within the harmonic oscillator
approximation. There are a total of 168 normal modes ($=56 \times
3$), six of which are translational and rotational modes. All of
these modes are vibrationally stable, which is also corroborated
by our frozen phonon calculations. Based on the normal coordinates
obtained from the vibrational analysis, we investigate which
modes are infrared (IR) and/or Raman active. IR absorption
intensity is proportional to the square of change of electric
dipole moment with respect to displacement along the normal
coordinates. Raman scattering intensity is proportional to
square of the change of polarizability with respect to displacement
along the normal coordinates. We calculate the IR absorption and
Raman scattering intensities following the method suggested in
Ref.~\cite{PORE96,WILS55,CARD82,MURP69}. First, we discuss the IR
spectra shown in Fig.~\ref{fig:IR}. Projection of the total IR
intensities onto each component in the monomer, indicates which
component contributes to the spectra at certain modes. Overall,
we find four strong IR intensities at 543~cm$^{-1}$,
1316~cm$^{-1}$, 1547~cm$^{-1}$, and 2942~cm$^{-1}$. For
low-frequency modes below 444~cm$^{-1}$ all parts of the monomer
contribute. At 543~cm$^{-1}$, the dominant contributions are from
O$^{2-}$ and Mn. The contributions of Mn and O$^{2-}$ are limited
to low-frequency modes (below 602~cm$^{-1}$). From 671~cm$^{-1}$
the contribution from pyridine rings becomes important. At
1316~cm$^{-1}$, 1547~cm$^{-1}$, 1563~cm$^{-1}$, and
2942~cm$^{-1}$ the contributions are from O$_2$CH and pyridine
rings. For high-frequency modes the vibrations of O$_2$CH and
pyridine rings provide dominant IR intensities. Second, we
discuss the Raman spectra shown in Fig.~\ref{fig:Raman}. The
Raman intensities are strong mostly at very low vibrational
frequencies, in contrast to the IR intensities that 
are strong at high frequencies. The strongest Raman
intensity is at 21~cm$^{-1}$, and the next largest peak is at
35~cm$^{-1}$. These modes are mainly due to vibrations of
pyridine rings as well as small contributions from Cl, Mn, and
O$_2$CH. At 76~cm$^{-1}$ and 96~cm$^{-1}$ the dominant
contributions are from Cl and smaller ones from pyridine,
O$_2$CH, and Mn. At 111~cm$^{-1}$, 168~cm$^{-1}$, and
212~cm$^{-1}$, the contributions are from all components except
for O$^{2-}$. At 514~cm$^{-1}$, 543~cm$^{-1}$, and 602~cm$^{-1}$
the contributions are from O$^{2-}$ and Mn only. At
1019~cm$^{-1}$, only pyridine rings contribute, and at
2942~cm$^{-1}$ (not shown) O$_2$CH and pyridine rings contribute.
At 3115~cm$^{-1}$, 3132~cm$^{-1}$, and 3154~cm$^{-1}$ (not shown)
only pyridine rings contribute. Compared to the IR spectra, the
contributions of Cl atoms are noticeable for low-frequency modes
and those of O$^{2-}$ appear at between 514~cm$^{-1}$ and
602~cm$^{-1}$ only. For high-frequency modes, the contributions
are from pyridine and O$_2$CH only.

\section{Conclusion}

We have found optimized geometries for the Mn$_4$ monomer
and dimer using DFT, and calculated exchange interactions between
Mn spins within the monomer and magnetic anisotropy barriers
for the ground state $S=9/2$ and the low-energy spin excitations.
We also have calculated the binding energy and the monomer-monomer
exchange interaction and magnetic dipole-dipole interaction
within a dimer. Most of our DFT calculations are in good agreement
with experiment, although the calculated intermolecular/intramolecular
exchange interactions are somewhat overestimated compared to the
experimental values as is common for DFT calculations. 
Our vibrational analysis on the hydrogenated
EDC monomer rules out a possibility of breaking
the threefold symmetry by stable soft vibrational modes in the
SMM Mn$_4$. The calculated IR and Raman spectra can be
checked by experiments.

\section*{Acknowledgments}
KP was funded by W. M. Keck Foundation, MRP and NB were supported
in part by ONR and NRL. Computer time was provided by the DoD HPCMPO 
at the ERDC MSRC.

\clearpage
\begin{table}
\begin{center}
\caption{Energy gaps between the highest
occupied molecular orbital (HOMO) and the lowest unoccupied
molecular orbital (LUMO) for majority and minority spins for
the SMM Mn$_4$ monomer.}
\label{table:1}
\begin{tabular}{|c|c|}\hline
maj HOMO - maj LUMO & min HOMO - min LUMO \\ \hline
 1.02 eV & 2.42 eV \\ \hline
min LUMO - maj HOMO & maj LUMO - min HOMO \\ \hline
1.17 eV & 2.28 eV \\ \hline
\end{tabular}
\end{center}
\end{table}

\begin{table}
\begin{center}
\caption{Eigenvalues of the $z$ component of the total spin operator,
$M_s$, spin configurations, their Ising exchange energies,
calculated energies relative to the ground state energy, $\Delta E$,
differences between the calculated and the least-square-fitted energies,
and calculated magnetic anisotropy barriers (MAE) per monomer for
the ground state and the low-energy spin excitations. For the spin
configurations, the first three symbols are for the three Mn$^{3+}$
spins ("$+$" denotes $M_z=+2$) and the fourth one is for the Mn$^{4+}$ spin
("$+$" denotes $M_z=+3/2$). Here $E_0$ is the background energy,
$J_1$ and $J_2$ are the intramolecular exchange constants
indicated in Fig.~\ref{fig:spin_conf}. These calculations were
performed for a computationally determined conformer (CDC).}
\label{table:2}
\begin{tabular}{|c|c|c|c|c|c|}\hline
$M_s$ & spin conf. & Ising energy & $\Delta E$ (eV) &
$E^{\mathrm{DFT}} - E^{\mathrm{LSF}}$(eV) & MAE/monomer (K) \\ \hline
9/2  & $+$ $+$ $+$ $-$ & $E_0 + 12 J_1 - 9 J_2$ & 0 & $-2.7 \times 10^{-4}$ & 11.3 \\ \hline
1/2  & $+$ $+$ $-$ $-$ & $E_0 - 4 J_1 - 3 J_2$ & 0.14 & $8 \times 10^{-4}$ & 11.4 \\ \hline
7/2  & $+$ $+$ $-$ $+$ & $E_0 - 4 J_1 + 3 J_2$ & 0.217 & $-8 \times 10^{-4}$ & 11.8 \\ \hline
15/2 & $+$ $+$ $+$ $+$ & $E_0 + 12 J_1 + 9 J_2$ & 0.236 & $2.7 \times 10^{-4}$ & 10.9 \\ \hline
\end{tabular}
\end{center}
\end{table}

\begin{table}
\begin{center}
\caption{Calculated intramolecular exchange constants,
energy gap between the ground-state $S=9/2$ and the first excited
doubly degenerate $S=7/2$ manifold, $E(S=7/2)$,
and energy gap between the ground-state and the second-excited
$S=11/2$ manifold, $E(S=11/2)$, in comparison with
the experimental results.\cite{HEND92,WERN02-NAT}
The negative sign means ferromagnetically coupled. DFT(1) denotes our
calculated values, while DFT(2) denotes values obtained by reducing
our calculated values by half. The last two columns are the
binding energy and intermolecular exchange constant within a CDC dimer.}
\label{table:3}
\begin{tabular}{|c|c|c|c|c||c|c|} \hline
 & $J_1$(K) & $J_2$(K) & $E(S=\frac{7}{2})$(K) &
$E(S=\frac{11}{2})$ (K) & Binding energy (eV) &
$J_{\mathrm{inter}}(K)$ \\ \hline 
DFT(1) & $-44$ & 152 &  490 & 836 & 0.16 & 0.24 \\ \hline 
DFT(2) & $-22$ & 76  &  245 & 418 & &           \\ \hline 
Exp    & $-25$\cite{HEND92} & 60\cite{HEND92} &
238\cite{HEND92} & 330\cite{HEND92} &  & 0.1\cite{WERN02-NAT} \\\hline
\end{tabular}
\end{center}
\end{table}

\end{document}